\documentclass[twocolumn, switch]{article} 

\usepackage{preprint}

\usepackage{amsmath, amsthm, amssymb, amsfonts}

\usepackage[numbers,square]{natbib}
\usepackage[utf8]{inputenc}	
\usepackage[T1]{fontenc}	
\usepackage{xcolor}		
\usepackage[colorlinks = true,
            linkcolor = black,
            urlcolor  = blue,
            citecolor = green,
            anchorcolor = black]{hyperref}	
\usepackage{booktabs} 		
\usepackage{nicefrac}		
\usepackage{microtype}		
\usepackage{lineno}		
\usepackage{float}			
\usepackage{cuted}

\usepackage{lipsum}		

\usepackage{newfloat}
\DeclareFloatingEnvironment[name={Supplementary Figure}]{suppfigure}
\usepackage{sidecap}
\sidecaptionvpos{figure}{c}

\usepackage{titlesec}
\titlespacing\section{0pt}{12pt plus 3pt minus 3pt}{1pt plus 1pt minus 1pt}
\titlespacing\subsection{0pt}{10pt plus 3pt minus 3pt}{1pt plus 1pt minus 1pt}
\titlespacing\subsubsection{0pt}{8pt plus 3pt minus 3pt}{1pt plus 1pt minus 1pt}

\usepackage{tikz,xcolor,hyperref}

\definecolor{lime}{HTML}{A6CE39}
\DeclareRobustCommand{\orcidicon}{
	\begin{tikzpicture}
	\draw[lime, fill=lime] (0,0) 
	circle [radius=0.16] 
	node[white] {{\fontfamily{qag}\selectfont \tiny ID}};
	\draw[white, fill=white] (-0.0625,0.095) 
	circle [radius=0.007];
	\end{tikzpicture}
	\hspace{-2mm}
}
\foreach \x in {A, ..., Z}{\expandafter\xdef\csname orcid\x\endcsname{\noexpand\href{https://orcid.org/\csname orcidauthor\x\endcsname}
			{\noexpand\orcidicon}}
}

\title{On-Chip Interferometric Excitation of an Infinity-Loop Microresonator}

\usepackage{authblk}

\author[1,*]{Davide Olivieri\orcidA{}}
\author[1]{B\"ulent Aslan\orcidB{}}
\author[1]{Stefano Biasi\orcidC{}}
\author[1,2]{Riccardo Franchi\orcidD{}}
\author[1]{Lorenzo Pavesi\orcidE{}}

\affil[1]{Nanoscience Laboratory, Department of Physics, University of Trento, 38123 Povo, Italy}
\affil[2]{Present address: Nanomaterials \& Nanostructure Optics, Department of Electrical and Computer Engineering, Boston University, Boston, MA, 02215, USA}
\affil[*]{Correspondence: \texttt{davide.olivieri@unitn.it}}

\begin{document}

\twocolumn[ 
  \begin{@twocolumnfalse} 
  
\maketitle

\begin{abstract}

Integrated photonics is a powerful platform for exploring Hermitian and non-Hermitian physics. Beyond device geometry, controlling how resonators are driven is crucial to access and tailor their modes. Coherent excitation via multiple input ports (interferometric excitation) enables such control, but its accurate description requires extending standard temporal coupled-mode theory to include interference between excitation pathways. Experimental realizations have so far been limited by phase-unstable, off-chip interferometers. Here we implement a fully integrated, phase-stable interferometric excitation scheme for an infinity-loop microresonator, an established structure operating on an exceptional surface, and use it to test the extended theory. Phase-dependent measurements in the linear and thermo-optic nonlinear regimes show that the relative phase between inputs governs the intracavity energy distribution, enabling up to a twofold increase of the circulating power compared to single-port excitation. This integrated platform enables reproducible studies of phase-dependent effects and coherent-control schemes in non-Hermitian photonic devices.

\end{abstract}
\vspace{0.35cm}

  \end{@twocolumnfalse} 
] 



In recent years, integrated photonics has become a powerful platform for exploring the physics of Hermitian and non-Hermitian systems \cite{El2018Non}. Photonic implementations include single microring resonators \cite{biasi2019hermitian}, coupled-resonator arrays \cite{li2025higherorder} and devices with engineered modal interactions \cite{lee2025chiral, APL_ILMR, calabrese2020unidirectional}. To model these structures, the temporal coupled mode theory (TCMT) \cite{Haus-TCMT,Christopoulos-TCMT} provides an intuitive framework describing the temporal evolution of resonant modes and their coupling to the external environment through an effective (non-)Hermitian Hamiltonian. This formalism is widely used for the design of advanced photonic systems which show enhanced optical sensing \cite{chen2017exceptional, carlo2022nonhermitian, franchi_controlled_2025,PhysRevA.93.033809,PhysRevLett.122.153902}, unidirectional reflection \cite{calabrese2020unidirectional}, chiral light transport \cite{lee2023chiral,bo2016chiral, chen2024electrically} and topological properties \cite{hafezi2013imaging, dai2024programmable}.

Beyond the choice of geometry and material loss, the way a microresonator is driven plays a central role in determining which modes are excited and how their spectra appear. 

Double injection schemes in microring resonators, in which either clockwise or counterclockwise mode is excited from two different coupling points, have been widely explored for shaping spectral responses by exploiting interference between two simultaneously injected fields\cite{cohen_response_2018,sun_compact_2023,hu_silicon_2023}. 
In these configurations, the interference occurs between excitation pathways associated with the same propagation direction. On the other hand, interferometric excitation relies on the simultaneous coherent excitation of both counter-propagating modes of a microresonator from a single coupling point\cite{Biasi-interferometric, biasi_interferometric_t_2022, FranchiPHD}. 
This approach leverages the interference between the two modal components through coherent control of the input amplitudes and phases, providing access to both the real and imaginary parts of the modal eigenvalues. However, the standard TCMT formulation, which is sufficient for single-point excitation, is generally not adequate for describing such multi-point coherent drivings. As recently emphasized by the theoretical work of Simonson et al.\ \cite{Simonson-destrInterf}, accurately capturing interference effects from multiple excitation pathways requires the inclusion of an additional phase term in the TCMT coupling matrix. At the same time, existing experimental implementations of interferometric excitation typically rely on off-chip fiber-based setups \cite{Biasi-interferometric, FranchiPHD}, where phase drifts caused by fiber relaxation and thermal fluctuations limit stability, reproducibility and the possibility of precise phase-dependent studies of complex resonant structures.

Here, we address these two limitations by combining an integrated non-Hermitian microresonator with a fully on-chip interferometric excitation scheme. We consider an infinity-loop microresonator (ILMR), a photonic structure capable of achieving either low- or high-contrast unidirectional reflection as well as a variety of spectral shapes, while always operating on an exceptional surface \cite{APL_ILMR, FranchiPHD}. The ILMR implements an effective unidirectional coupling between counterpropagating modes, side-coupled to a bus waveguide and coherently driven from two opposite directions. The relative amplitudes and phases of the two inputs are controlled by integrated Mach--Zehnder interferometers and heaters, providing a phase-stable platform for multi-port excitation. Following the extended TCMT approach of Ref.~\cite{Simonson-destrInterf}, we introduce an explicit phase term in the coupling matrix to account for interference between distinct excitation paths and derive an effective non-Hermitian Hamiltonian for the ILMR under interferometric excitation. We then use this framework, in combination with transfer-matrix simulations and experiments on a silicon photonic chip, to perform a phase-dependent analysis of the ILMR response. In the linear regime, we show that including the additional phase term is necessary to reproduce the measured evolution of the modal spectrum as a function of the excitation phase. In the nonlinear regime, we exploit the thermo-optic nonlinearity of silicon to demonstrate that the same interferometric control can be used to adjust the intracavity intensity at fixed input power and to toggle between weakly nonlinear and thermally bistable operation.

\begin{figure}[th!]
\centering
\includegraphics[width=0.9\linewidth]{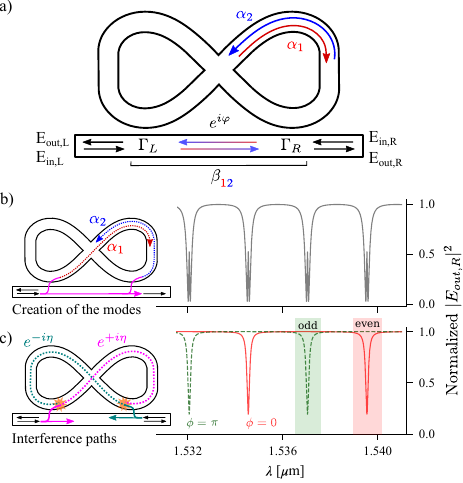}
\caption{\label{fig:fig1}Schematic description and behavior of the ILMR. (a) Sketch of the microresonator; all symbols are defined in the main text. (b) Generation of the two counterpropagating modes under left excitation, scheme and simulated spectral response. (c) Interferometric excitation: interference paths between the two coherent excitation fields. The simulated spectra distinguish between \textit{even} (solid red) and \textit{odd} (dashed green) resonances.}
\end{figure}

The ILMR, sketched in Fig.~\ref{fig:fig1}(a), is composed by an infinity-shaped microresonator with both lobes coupled to a straight bus waveguide. TCMT provides a powerful formalism to describe its non-Hermitian exceptional point-based operating regime \cite{APL_ILMR, Modeling_ILMR}. Specifically, the microresonator holds two counter-propagating modes, $\alpha_1$ and $\alpha_2$, represented, respectively, by the red and blue arrows in Fig.~\ref{fig:fig1}(a). The coupling between the infinity-shaped waveguide and the bus waveguide occurs in two different regions, and it is modeled by the rates $\Gamma_{\rm L}$ and $\Gamma_{\rm R}$, where the indexes L and R refer to the left and right sides of the microresonator. In case of a single sided excitation, both modes are excited simultaneously from the two coupling regions, as illustrated in Fig.~\ref{fig:fig1}(b). Due to the microresonator design, the only interaction between counter-propagating modes is mediated by the bus waveguide. Specifically, the energy transfer is possible only from mode $\alpha_1$ to $\alpha_2$ and is parametrized by the nonzero coupling rate $\beta_{12}$$=4e^{i\varphi}\sqrt{\Gamma_L\Gamma_R}$ (while no coupling occurs from $\alpha_2$ to $\alpha_1$; therefore, the coupling rate $\beta_{21}=0$). With reference to the symbols in Fig. \ref{fig:fig1}(a), the ILMR spectral response is described by a set of coupled first-order differential equations:

\begin{equation}\begin{split}
    i \frac{d}{dt} 
    \begin{bmatrix} 
        \alpha_2 \\ 
        \alpha_1 
    \end{bmatrix} 
    = 
    \begin{pmatrix} 
        \omega_0 - i \gamma_{\rm tot} & -i \beta_{12} \\ 
        0 & \omega_0 - i \gamma_{\rm tot} 
    \end{pmatrix} 
    \begin{bmatrix} 
        \alpha_2 \\ 
        \alpha_1 
    \end{bmatrix} 
    + \\ -
    \begin{pmatrix} 
        \sqrt{2 \Gamma_{\rm R}} e^{i \varphi} & \sqrt{2 \Gamma_{\rm L}} e^{i \varphi}e^{-i\eta} \\ 
        \sqrt{2 \Gamma_{\rm L}} & \sqrt{2 \Gamma_{\rm R}}e^{-i\eta} 
    \end{pmatrix} 
    \begin{bmatrix} 
        E_{\rm in,L} \\ 
        E_{\rm in,R}e^{i\phi} 
    \end{bmatrix}.
    \end{split}
    \label{eq:TCMT_modes}
\end{equation}

\noindent 
Here, $E_{\text{in,L/R}}$ are the amplitudes of the input fields at L (left) or R (right) ports, with relative phase $\phi$. $\gamma_{\rm tot}$ represents the total loss rate, including both \textit{intrinsic} ($\gamma$) and \textit{extrinsic} losses ($\gamma_{\rm tot} = \Gamma_{\rm L} + \Gamma_{\rm R} + \gamma$). Additionally, $\varphi$ is the phase acquired by the optical modes and the input fields as they propagate through the bus waveguide between the two coupling points. $\eta$ is an additional phase introduced to account for the two-port excitation, as will be discussed below. The first matrix in Eq.~\ref{eq:TCMT_modes} is the system's effective Hamiltonian $H$, while the second matrix describes the coupling between the resonant modes and the two input fields. The output fields, $E_{\text{out,L/R}}$, are related to both the input fields and the intracavity modes of the ILMR by the following relation:

\begin{equation}
    \begin{split}
    \begin{bmatrix} 
        E_{\rm out,R} \\ 
        E_{\rm out,L} 
    \end{bmatrix} 
    = 
    e^{i\varphi}
    \begin{bmatrix} 
        E_{\rm in,L} \\ 
        E_{\rm in,R}e^{i\phi} 
    \end{bmatrix} 
     +i 
    \begin{pmatrix} 
        \sqrt{2 \Gamma_{\rm R}} & \sqrt{2 \Gamma_{\rm L}} e^{i \varphi}\\ 
        \sqrt{2 \Gamma_{\rm L}}e^{+i\eta}  & \sqrt{2 \Gamma_{\rm R}} e^{i \varphi}e^{+i\eta} 
    \end{pmatrix} 
        \begin{bmatrix} 
        \alpha_2 \\ 
        \alpha_1 
    \end{bmatrix} .
    \end{split}
    \label{eq:TCMT_fields}
\end{equation} 
\noindent
The steady-state solution of Eq.~\ref{eq:TCMT_modes} provides an explicit expression for the two modes $\alpha_1$ and $\alpha_2$\cite{APL_ILMR,FranchiPHD}. In the case of a symmetric ILMR ($\Gamma_{\rm L}=\Gamma_{\rm R}=\Gamma$), this leads to: 
\begin{equation}
    \begin{aligned}
        \alpha_2 &= \frac{i\sqrt{2\Gamma}e^{i\varphi}}{-i\Delta\omega + \gamma_{\rm tot}} \left( 1- \frac{4\Gamma}{-i\Delta\omega + \gamma_{\rm tot}} \right)\left( E_{\rm in,L} + E_{\rm in,R}e^{i(\phi-\eta)} \right), \\
        \alpha_1 &= \frac{i\sqrt{2\Gamma}}{-i\Delta\omega + \gamma_{\rm tot}} \left( E_{\rm in,L} + E_{\rm in,R}e^{i(\phi-\eta)} \right),
    \end{aligned}
    \label{eq:a1a2}
\end{equation}
\noindent
where $\Delta\omega = \omega-\omega_0$. Together with the input-output relation given in Eq.~\ref{eq:TCMT_fields}, these solutions fully describe the system’s response under single-sided excitation \cite{APL_ILMR}. However, when the additional $\eta$ phase is omitted, they fail to capture the physics of coherent double-sided excitation. In fact, the coexistence of multiple excitation pathways gives rise to an additional, phase-dependent interference effect \cite{Simonson-destrInterf}. Therefore, we incorporate an additional phase $\eta$ to account for the path difference in mode excitation between the two input ports, defined with respect to the excitation point associated with $E_{\rm in,L}$ (see Fig.~\ref{fig:fig1}(b)). Specifically, $\eta$ denotes the phase accumulated along one half of the ILMR (dashed pink lines in Fig.~\ref{fig:fig1}(c)), represented by the phasor $e^{+i\eta}$. To satisfy the resonance condition, the phase accumulated along the complementary path (dashed green lines) must be $2\pi-\eta$, corresponding to the phasor $e^{-i\eta}$, such that the total round-trip phase is $2\pi$ (or an integer multiple of $2\pi$). As shown in Fig.~\ref{fig:fig1}(c), when both inputs are active, interference occurs between the field entering the nearest coupling region (solid lines) and the one propagating through one lobe of the ILMR (dashed lines). For instance, light entering from the left (pink solid line) interferes with the field coupled from the right side after traversing half the ILMR (dashed green lines). In the case of a symmetric microresonator, this relative phase is
\begin{equation}
\eta = \frac{2\pi n_\mathrm{eff}}{\lambda} \frac{L}{2},
\label{eq:eta}
\end{equation}
where $L$ is the ILMR perimeter, $\lambda$ is the wavelength of the injected light and $n_{\mathrm{eff}}$ is the 
effective refractive index of the guided mode. When the resonance condition $\frac{2\pi n_\mathrm{eff}}{\lambda} L = 2\pi m$ ($m \in \mathbb{Z}$) is considered, the phase reduces to $\eta = \pi m$, allowing distinction between \textit{even} and \textit{odd} modal orders. 

The phase $\eta$ directly impacts the modes excitation. Under the assumption of a symmetric ILMR, the system in Eq.~\ref{eq:TCMT_modes} reduces to

\begin{equation}
i \frac{d}{dt} 
    \begin{bmatrix} 
        \alpha_2 \\ 
        \alpha_1 
    \end{bmatrix} 
    = 
    H 
    \begin{bmatrix} 
        \alpha_2 \\ 
        \alpha_1 
    \end{bmatrix} 
    - \sqrt{2\Gamma}
    \begin{pmatrix}
        e^{i\varphi}(E_{\rm in,L} + E_{\rm in,R}e^{+i(\phi -\eta)}) \\
        (E_{\rm in,L} + E_{\rm in,R}e^{+i(\phi - \eta)})
    \end{pmatrix}.
    \label{eq:TCMT_simpl}
\end{equation}

\noindent
For a balanced interferometric excitation $|E_{\rm in,L}| = |E_{\rm in,R}|$, the phase difference $(\phi - \eta)$ plays a crucial role. Depending on the modal order, the phasor $e^{i\eta}$ equals $\pm 1$ and a different relative phase between the two input fields is required to suppress the creation of the two modes. Specifically, when $e^{i(\phi - \eta)} = -1$, the second vector in Eq.~\ref{eq:TCMT_simpl} becomes the zero vector, as the excitation fields undergo complete destructive interference within the microresonator. This leads to a \textit{flat} transmission, with the microresonator exhibiting no features, equivalent to an off-resonant condition.
When $e^{i(\phi - \eta)} = +1$, constructive interference occurs, and the modal excitation is not suppressed. Therefore, interferometric excitation exploits the ILMR symmetry to selectively excite \textit{even} or \textit{odd} resonances through the relative phase $\phi$ tuning. This behavior arises from the coherent driving through two distinct coupling points and it is not explicitly related to the exceptional point nature of the system.

We have verified the TCMT predictions through the Transfer Matrix Method. The ILMR was modeled by using the designed parameters: a $224\,\mu\textit{m}$-long infinity-loop with a $48\,\mu\textit{m}$ separation between the two coupling regions, both characterized by an identical transmission coefficient $t=0.946$. Linear propagation losses were set to $2\,\textit{dB/cm}$, and the group index for the guided mode is $n_{g} = 4.22$. The resulting spectra are reported in Fig.~\ref{fig:fig1} (b) and (c). Under single sided-excitation [Fig.~\ref{fig:fig1}(b)], the ILMR displays the typical doublet response in transmission due to the excitation of both modes and their mutual interference in the bus waveguide. In contrast, balanced interferometric excitation [Fig.~\ref{fig:fig1}(c)] produces a qualitatively different behavior: (i) for $\phi=0$ (solid red curves), the free spectral range effectively doubles, responding only to the \textit{even} resonances (as the one highlighted in the red rectangle). For \textit{even}-order resonances, $\eta$ is an integer multiple of $2\pi$, requiring a relative input phase of $2m\pi$, with $m\in\mathbb{Z}$, for constructive excitation. (ii) for $\phi=(2m+1)\pi$ (green dashed curves), the \textit{odd} resonances emerge (green rectangle), while the \textit{even} ones  become flat. Due to the ILMR’s structural symmetry, the output spectra at both ports are identical for the phase relations considered; thus we only show $|E_{\rm out,R}|^2$ as the representative signal. These simulations provide a direct numerical confirmation of the phase-selective modal excitation predicted by the extended TCMT model.

\begin{figure}[th!]
\centering
\includegraphics[width=0.9\linewidth]{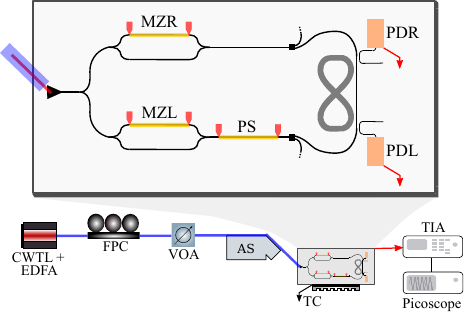}
\caption{\label{fig:setup}Sketch of the photonic integrated circuit and the experimental setup. Symbols are defined in the main text.}
\end{figure}

The theoretical and numerical results were subsequently validated experimentally. The experimental setup and the design of the photonic integrated circuit are schematically shown in Fig.~\ref{fig:setup}. The device was fabricated on a silicon-on-insulator platform using a CMOS-compatible process by AMF/Europractice within a multi-project wafer run. The ILMR curved sections have a radius of 15 $\mu$m, chosen to avoid bending losses. The optical source is a continuous wave tunable laser (CWTL, TUNICS T100S-HP from Yenista) with a fibered output. Higher powers, required to observe nonlinear effects, are obtained by inserting an erbium-doped fiber amplifier (EDFA) after the CWTL. A fiber polarization controller (FPC) and a micrometric alignment stage (AS) ensure proper TE polarization and optical coupling between the angled stripped fiber and the input grating coupler (black triangle in the figure). To maximize the coupling efficiency, a layer of liquid silicone is deposited on the surface as index-matching material. Once coupled, the input light is equally split into two arms that excite the ILMR from the opposite sides. Each arm contains an integrated Mach-Zehnder interferometer (MZL and MZR) used to control the excitation amplitudes $|E_{\rm in,L/R}|$. In addition, the left arm includes an integrated phase-shifter (PS) to adjust the relative phase $\phi$. Both the MZs and the PS employ 2-$\mu m$-wide Titanium Nitride (TiN) micro-heaters deposited on top of the waveguides, with a resistance of 1457 $\Omega$. Each micro-heater is independently driven by a current generator. In this architecture, the signal is then collected by two integrated germanium photodetectors (PDL and PDR in Fig.~\ref{fig:setup}) that are weakly coupled to the bus waveguide. Their electrical outputs are then amplified by a transimpedance amplifier (TIA) and recorded with a PC oscilloscope (PicoScope 4000). 

\begin{figure*}[th!]
\includegraphics[width=1\linewidth]{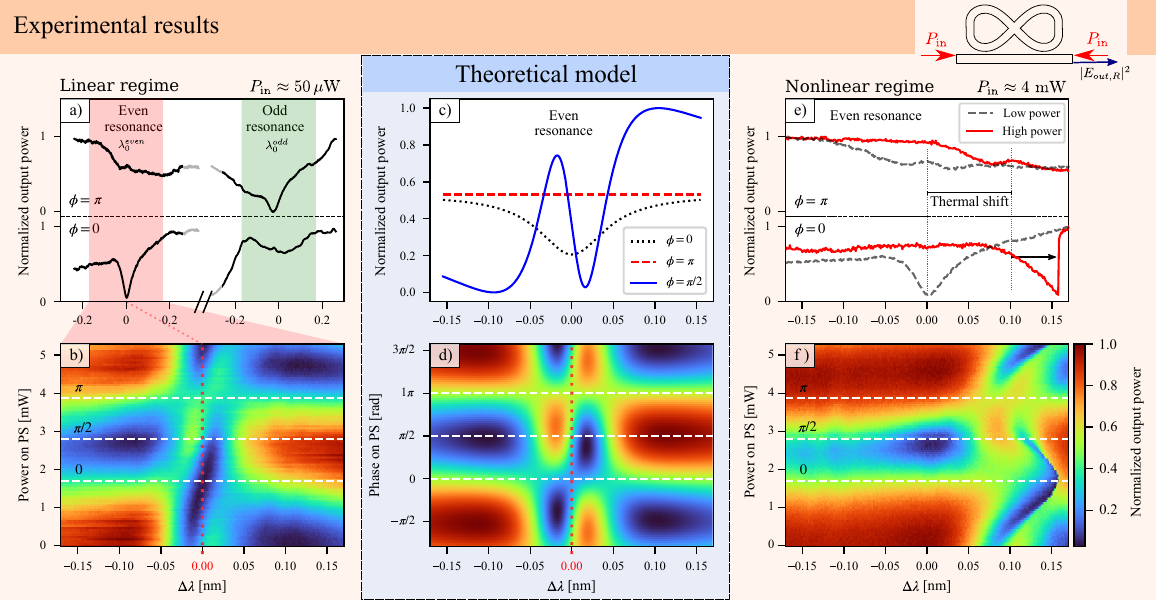}
\caption{\label{fig:TeoAndExp}Experimental measurements and theoretical results. All panels report the normalized output power $|E_{\rm out,R}|^2$ under balanced interferometric excitation (as shown in the upper right inset). (a) Measured low-power spectra for the \textit{even} resonance ($\lambda_0^{\rm even} = 1536.00$ nm, red region) and an \textit{odd} one ($\lambda_0^{\rm odd} = 1558.20$ nm, green region) under two phase conditions, $\phi = \{0,  \pi\}$. (b) corresponding experimental colormap showing $|E_{\rm out,R}|^2$ versus the power dissipated by the PS and the wavelength detuning $\Delta\lambda$, centered around $\lambda_0^{\rm even}$. (c) TCMT simulations for different phase values $\phi$ in the case of an \textit{even} resonance. (d) Theoretical colormap corresponding to (b). For these simulations, the parameters are $\Gamma = 17$ GHz, $\gamma=4$ GHz, $|E_{\rm in,L/R}|^2=1$ mW. (e) Comparison between low-power (black dashed lines) and high-power (red solid lines) spectra for the \textit{even} resonance in the constructive and destructive interference cases. (f) high-power colormap analogous to the low-power one in (b). In the colormaps, vertical red dotted lines indicate the zero wavelength detuning ($\Delta\lambda=0$) while the horizontal white dashed lines highlight the main phase values $\phi= \{0,  \pi/2 , \pi\}$. In the maps, the output powers are normalized to their global maximum in the displayed range to enable a direct comparison. The spectra in (a) and (c), instead, are normalized individually to their respective maxima within the plotted spectral range.}
\end{figure*}

This fully on-chip interferometric excitation preserves optical coherence, enabling precise characterization of the system response as a function of the phase difference $\phi$, minimizing external phase fluctuations. Experimental results are shown in Fig.~\ref{fig:TeoAndExp}, displaying $|E_{\rm out,R}|^2$ as indicated in the inset at the upper right. For reference, we define the resonance centered at $\lambda_0^{\rm even} = 1536.00$ nm as \textit{even} and consequently $\lambda_0^{\rm odd} = 1558.20$ nm as \textit{odd}. Fig.~\ref{fig:TeoAndExp}(a) shows the spectra at two phase values, emphasizing the $\lambda_0^{\rm even}$ and $\lambda_0^{\rm odd}$ resonances while omitting intermediate responses for clarity. It should be noted that the experimental spectra exhibit strong Fabry–Pérot-like oscillations due to the device architecture, where part of the optical power is routed to other on-chip structures (not discussed here) at the positions indicated by black squares in  Fig.~\ref{fig:setup}; these splitters produce strong back-reflections that distort the signal. Additionally, spurious reflections originating from the integrated Mach–Zehnder interferometers and the input grating are also present. While Fabry–Pérot-like oscillations can be effectively modeled in simpler geometries \cite{calabrese2020unidirectional}, in our case the response arises from the interplay of multiple cavities, precluding a straightforward analytical treatment. Nevertheless, the main phase conditions ($\phi = 0$ and $\phi = \pi$) were successfully extrapolated by identifying the characteristic singlet-like and flat responses corresponding to the absence of the microresonator-induced modulation on the oscillations. Intermediate phase values were then obtained by linear interpolation. As expected, when $\phi = 0$, the \textit{even} resonance exhibits a single dip, while the \textit{odd} one remains flat. After a $\pi$ phase shift, the behavior is inverted: the \textit{odd} resonance displays a singlet-like feature, while the \textit{even} one becomes flat. The overall phase dependence is visualized through the colormap in Fig.~\ref{fig:TeoAndExp}(b), where the color scale represents the measured optical power $|E_{\rm out,R}|^2$ as functions of the power dissipated by the phase-shifter, computed from the injected current using Joule's law, and the wavelength detuning $\Delta\lambda$, centered around $\lambda_0^{\rm even}$. When moving from $\phi =0$ to $\phi = \pi$, the maxima and minima of the response reverse, as evidenced by the alternating color pattern in the map. The results are reported only for an \textit{even} resonance, since \textit{odd} resonances differ only by a $\pi$ phase shift (see Eq.~\ref{eq:TCMT_simpl}).

The observed phase-dependent spectral response is further analyzed using the steady state solution of the TCMT. Utilizing the parameters $\Gamma = 17$ GHz and $\gamma=4$ GHz, extracted from the fitted single-sided excitation spectra. The validity of TCMT is justified since both the intrinsic and extrinsic losses remain much smaller than the Free Spectral Range (FSR $\approx$ 2$\pi$ 308 GHz), ensuring that different mode overlap can be neglected. The balanced interferometric response was calculated for an \textit{even} modal order ($e^{i\eta}=+1$) with equal input amplitudes. The theoretical spectra given in Fig.~\ref{fig:TeoAndExp}(c) confirm the measured behavior: when $\phi = \pi$ (red dashed line), the response is flat (at $\sim$ 0.5 due to normalization), conversely, at $\phi =0$ (black dotted line) has a Lorentzian-like dip. For intermediate phase values (e.g. $\phi = \pi/2$, blue solid line), the interference of the two modes in the bus waveguide creates an asymmetric spectrum with respect to $\lambda_0$. Furthermore, the complete phase dependence displayed in Fig.~\ref{fig:TeoAndExp}(d) captures the main features of the experimental results presented in Fig.~\ref{fig:TeoAndExp}(b). Note that the results are computed within the same spectral range of the map in Fig.~\ref{fig:TeoAndExp}(b) and normalized to the global maximum of the map in Fig.~\ref{fig:TeoAndExp}(d).

To demonstrate how the interferometric phase $\phi$ control governs the intracavity energy, we also investigated the ILMR response at high input powers (around $4$ mW on each side of the bus waveguide), sufficient to trigger nonlinear effects. 
At 1.55 $\mu$m, the nonlinear response of silicon microresonators is dominated by two photons absorption-induced free carriers and the resulting thermal effects, rather than by the Kerr nonlinearity\cite{borghi_nonlinear_2017}. Therefore, we exploit the thermo-optic redshift and the associated bistable transition as probes of the intracavity energy. 
In all scans, the wavelength is swept from shorter to longer values at a constant rate of 5 nm/s.
Fig.~\ref{fig:TeoAndExp}(e) compares the low-power (i.e., linear regime; black dashed line) and high-power (i.e., nonlinear regime; red solid line) spectra for the same reference resonance centered at $\lambda_0^{\rm even}$. When $\phi = \pi$, destructive interference suppresses light circulation within the microresonator, and the only observable effect is a global thermal redshift ($\approx100$ pm) caused by laser-induced heating, as highlighted in the plot. Conversely, under constructive interference ($\phi = 0$), the intracavity intensity is maximized, leading to a resonance redshift and the characteristic hysteretic jump of optical bistability \cite{Boyd2008, RamiroManzano2013, Bernard2017}. The corresponding colormap in Fig.~\ref{fig:TeoAndExp}(f) clearly shows the overall thermal redshift and the emergence of the nonlinear bistable response near $\phi = 0$. The bistability amplitude gradually decreases as $\phi$ approaches $\pi$, where the intracavity power, thus the nonlinear contribution, is strongly reduced. These results demonstrate that we can control the internal optical power by tuning the relative phase $\phi$, effectively switching between distinct linear and nonlinear regimes.

\begin{figure*}[h]
    \includegraphics[width=1\linewidth]{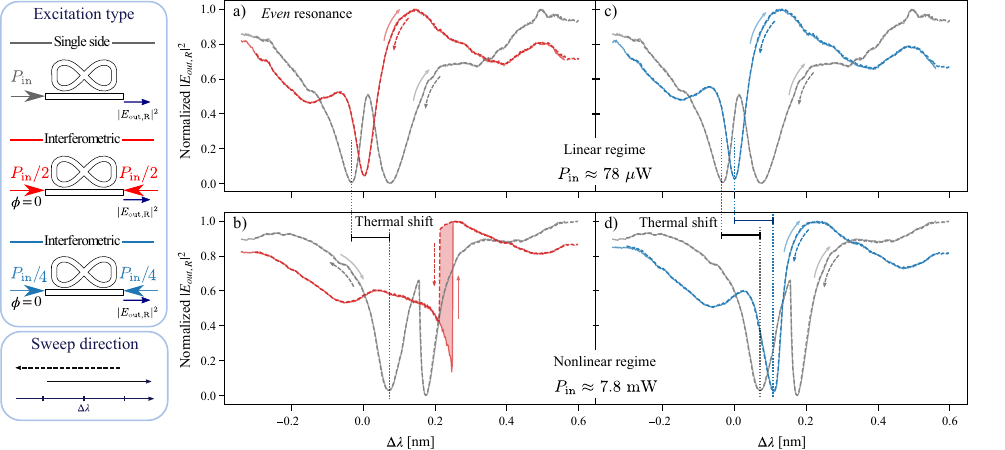}
    \caption{\label{fig:fig4} Experimental comparison between single-sided and interferometric excitation in the linear (a,c) and nonlinear (b,d) regimes. Transmission spectra from the right port $|E_{\rm out,R}|^2$ are recorded by scanning from shorter-to-longer wavelengths (solid lines) and backward (dashed lines). In all the cases, the PS is set to $\phi=0$, ensuring constructive interference requirement for the given \textit{even} resonance ($\lambda_{0}^{even} = 1557.44$ nm). Panel on the left: the different excitation configurations. Single sided excitation from the left port with an input power $P_{\rm in}$ (gray lines in a,b,c,d). Interferometric excitation from the left and right ports with an input power $P_{\rm in}/2$ (red lines in a,b). Interferometric excitation with an input power $P_{\rm in}/4$ (blue lines in c,d).}
\end{figure*}

This is a distinctive feature of the interferometric excitation by which constructive interference leads to a substantial buildup of intracavity power. 
Let us define the total intracavity power as $P_{\rm int}=|\alpha_1|^2+|\alpha_2|^2$.
Substituting Eq.~\ref{eq:a1a2} into this definition:
\begin{strip}
\begin{equation}
    P_{\rm int} = \frac{2\Gamma}{|-i\Delta\omega +\gamma_{\rm tot}|^2}\left( 1 + \left| 1 - \frac{4\Gamma}{-i\Delta\omega +\gamma_{\rm tot}}\right|^2\right) \left|  E_{\rm in,L} + E_{\rm in,R}e^{i(\phi-\eta)}   \right|^2.
\end{equation}
\end{strip}
\noindent The first term is common to both excitation configurations and depends only on the intrinsic ILMR parameters and detuning. Consequently, the difference in the intracavity power between single-sided and interferometric excitation arises solely from the last interference term. Specifically, when the phasor $e^{i(\phi-\eta)}=+1$, $P_{\rm int}$ scales with the square of the total excitation field $|E_{\rm in,L} + E_{\rm in,R}|^2$.
For comparison, under single-sided excitation, where $P_{\rm in} = |E_{\rm in,L}|^2$ and $|E_{\rm in,R}|^2 = 0$, the intracavity power scales linearly as $P_{\rm int} \propto P_{\rm in}$. 
In contrast, for balanced excitation of the same total input power ($|E_{\rm in,L}|^2 = |E_{\rm in,R}|^2 = P_{\rm in}/2$), the intracavity power scales as $P_{\rm int}\propto 2P_{\rm in}$. 
Under constructive interference, this facilitates a twofold enhancement of the intracavity power buildup compared to single-port excitation. 
An experimental comparison between these two conditions for both linear and nonlinear regimes are reported in Fig.~\ref{fig:fig4} (a) and (b). 
To probe optical bistability, the spectra are acquired by scanning the wavelength in both directions, from shorter to longer wavelengths (solid lines) and vice versa (dashed lines), as illustrated in the lower part of the panel on the left.
Each acquisition is performed point by point with a wavelength step of 2 pm and an average scan speed of 3.8 pm/s.
Single-sided excitation is achieved by setting MZL to its maximum transmission state and MZR to its minimum (no-transmission) state, thereby delivering the total input power $P_{\rm in}$ to the left ILMR port (see the upper part of the panel). 
For interferometric excitation, both MZL and MZR are set to their half-transmission states, ensuring a balanced input power of $P_{\rm in}/2$ at each microresonator port (see the second configuration in the panel). Since we have chosen an even resonance for this demonstration, the PS is set to $\phi=0$ to satisfy the condition for constructive interference.
The spectra obtained under single-sided excitation (gray lines in Fig.~\ref{fig:fig4} (a) and (b)) show only a minor nonlinear deformation and a thermal redshift, with no evidence of optical bistability. 
In contrast, under interferometric excitation (red lines), the spectra exhibit a stronger nonlinear shift and a pronounced bistable loop, highlighted by the red-shaded area. This is consistent with enhanced optical power circulation within the ILMR compared to the single-sided configuration.
Note that the input power $P_{\rm in}$ in Fig.~\ref{fig:fig4} (b) is comparable to that used in Fig.~\ref{fig:TeoAndExp}(e), and consequently the observed thermal redshifts in the two cases are similar. 
To further confirm the twofold enhancement of the intracavity energy, Fig.~\ref{fig:fig4}(c) and (d) compare the single-sided excitation with an interferometric excitation obtained by setting both MZs to 25\% transmission, thereby injecting an input power of $P_{\rm in}/4$ at each port (see the third configuration in the panel). In this case, the total input power is half of that used under single-sided excitation. Owing to the twofold enhancement of the intracavity power provided by the constructive interference, the same intracavity energy is therefore expected as in the case of single-side excitation. Accordingly, the interferometric excitation should exhibit a thermal response similar to that of the single-sided case. Indeed, the interferometric single resonance dip (blue lines) shows no bistable behavior and exhibits only an overall thermal redshift similar to the one of the single-sided excitation.

In conclusion, we have realized a phase-stable, fully integrated interferometric excitation scheme for an on-chip infinity-loop microresonator and used it to investigate its phase-dependence on the input excitations. Within a temporal coupled-mode theory that includes an additional phase term in the coupling matrix, we obtain a compact model that captures the effect of multi-path interference and reproduces the measured phase-dependent spectra. Earlier models, which do not include phase dependence, represent a special case of our more general modal framework: setting $e^{i\eta}=1$ recovers the previous results \cite{APL_ILMR,Modeling_ILMR}. Our approach accounts for arbitrary $\eta$, encompassing both \textit{even} and \textit{odd} resonances as well as the full range of interferometric effects. The experiments show that the relative phase between the two input ports controls the excitation of different resonances in the linear regime and, at fixed input power, tunes the intracavity intensity and the onset of thermo-optic bistability in the nonlinear regime. As a direct consequence, interferometric control enables a twofold enhancement of the intracavity power compared to conventional single-port excitation. This compact and scalable architecture provides a convenient testbed for interferometric excitation and coherent control in integrated photonics and can serve as a building block for more complex non-Hermitian and phase-programmable photonic circuits.

\footnotesize
\section*{Funding} This project has received funding from the European Research Council (ERC) under the European Union’s Horizon 2020 research and innovation programme (grant agreement No. 788793, BACKUP).

\footnotesize
\section*{Acknowledgements}
The authors acknowledge the insightful discussions with Mr. Diego Piciocchi on the response of the infinity-loop microresonator under interferometric excitation.  

\footnotesize
\section*{Disclosures} The authors declare no conflicts of interest.

\footnotesize
\section*{Data availability} Data underlying the results presented in this paper are not publicly available at this time but may be obtained from the authors upon reasonable request.



\nocite{*}
\bibliography{References}


\end{document}